\documentclass[aip,cha,amsmath,amssymb,reprint]{revtex4-1}
\usepackage{color}
\usepackage{graphicx}
\usepackage{dcolumn}
\usepackage{bm}
\usepackage[mathlines]{lineno}
\usepackage{amsfonts}
\usepackage{amsmath}
\usepackage{sidecap}
\usepackage{float}
\usepackage{appendix}
\usepackage{parskip}
\usepackage{grffile}

\usepackage{color}
\usepackage{hyperref}
\usepackage{hhline}
\usepackage{mathtools}
\usepackage{comment}
\usepackage[normalem]{ulem}
\makeatletter
\def\@eqnnum{{\normalsize \normalcolor (\theequation)}}
 \makeatother
\begin{document}

\title{Finite size effect in Kuramoto oscillators with inertia on simplicial complex}

\author{Manuel Lourenço}\thanks{These authors contributed equally.}
\affiliation{Fraunhofer Institute for Algorithms and Scientific Computing, Sankt-Augustin-53757, Germany}
\author{Abhishek Sharma}\thanks{These authors contributed equally.}
\affiliation{Complex Systems Lab, Department of Physics, Indian Institute of Technology Indore, Khandwa Road, Simrol, Indore-453552, India}

\author{Priyanka Rajwani}
\affiliation{Complex Systems Lab, Department of Physics, Indian Institute of Technology Indore, Khandwa Road, Simrol, Indore-453552, India}
\author{Erick Alejandro Madrigal Solis}
\affiliation{Fraunhofer Institute for Algorithms and Scientific Computing, Sankt-Augustin, Germany}
\affiliation{Dresden University of Technology, Dresden-01069, Germany}
\author{Mehrnaz Anvari}\email{mehrnaz.anvari@scai.fraunhofer.de}
\affiliation{Fraunhofer Institute for Algorithms and Scientific Computing, Sankt-Augustin, Germany}
\affiliation{Potsdam Institute for Climate Impact Research, Potsdam, Germany}

\author{Sarika Jalan}\email{Corresponding Author: sarikajalan9@gmail.com}
\affiliation{Complex Systems Lab, Department of Physics, Indian Institute of Technology Indore, Khandwa Road, Simrol, Indore-453552, India}

 \begin{abstract}
We investigate the finite-size effects on the dynamical evolution of the Kuramoto model with inertia coupled through triadic interactions. Our findings reveal that fluctuations resulting from the finite size drive the system toward a synchronized state at finite coupling, which contrasts with the analytical predictions {in thermodynamic limit} made for the same system. Building on the analytical calculations performed at the thermodynamic limit, we identify the origin of the synchronization transition that arises because of the finite size. We discover a power-law relationship between the network size and the critical coupling at which the first-order transition to synchronization occurs. Additionally, as inertia increases, there is a significant shift in the critical coupling toward higher values, indicating that inertia counteracts the effects caused by finite size.
%
\end{abstract}
  
\maketitle

\begin{quotation}
  Kuramoto oscillators with inertia serve as a fundamental model for studying the dynamic behaviors of various complex systems, including diluted networks found in Josephson junctions and power grids. This model is widely used to understand the origins of coherence in interacting dynamical units. Although coupled Kuramoto oscillators have been extensively studied regarding pairwise interactions, research on models that incorporate inertia and higher-order couplings remains limited. In this study, we demonstrate that analytical predictions made using a model system's thermodynamic limit may not accurately reflect the behaviors of real-world systems that are finite in size. Since nearly all real-world systems are finite, it is essential to investigate how finite size impacts dynamic behaviors. We specifically examine coupled Kuramoto oscillators with inertia and higher-order interactions, such as 2-simplices. Our findings reveal that finite-size effects can induce a transition to synchronization—an outcome that is not predicted by analytical models of the same system.
\end{quotation}

\paragraph*{\bf{Introduction:}}
Synchronization over time is an emerging phenomenon in nature, primarily arising from the interactions among dynamical units. Examples of synchronization in real-world systems include synchronous flashing of fireflies \cite{buck1988synchronous}, coordinated chirping of crickets \cite{walker1969acoustic}, collective flocking of birds \cite{attanasi2015emergence}, and brain function \cite{osipov2007synchronization}. 
Such coherent behavior is often driven by a phase transition that transforms the system from a disordered state to a synchronized one. Coupled phase oscillators proposed by Kuramoto \cite{kuramoto1975self} is one of the most popular models to study the origin of synchronization phase transitions in interacting dynamical units. In this model, each oscillator rotates with its intrinsic frequency, $\omega$, and interacts with other oscillators through a nonlinear (sinusoidal) coupling function. It has been shown that Kuramoto oscillators with pairwise coupling yield a smooth transition from an incoherent state to a synchronized state through a supercritical bifurcation as the coupling strength increases \cite{STROGATZ20001}. 
Furthermore, to model the synchronized flashing in Pteroptyx malaccae, Ermentrout utilized the Kuramoto oscillator model \cite{Ermentrout}, in which Tanaka proposed an inertia term \cite{PhysRevLett.78.2104tanaka,TANAKA1997279}.
The addition of the inertia term has been shown to facilitate the synchronization transition at a higher critical coupling value \cite{TANAKA1997279}. Later, this second-order Kuramoto model was explored \cite{PhysRevLett.110.218701Kurth} for diluted networks \cite{PhysRevE.71.016215santhanam}, in Josephson junctions, and power grids \cite{10.1063/1.4967850Grzybowski, doi:10.1073/pnas.1212134110bullo, PhysRevLett.109.064101MArcRhoden, RevModPhys.94.015005kurthMarc}.
 
The last few years have seen a substantial increase in research on higher-order interactions. Incorporating higher-order interactions into the classical Kuramoto model (without inertia) has been shown to lead to multi-stable states corresponding to different initial conditions \cite{skardal2019abrupt}. The importance of higher-order interactions has been elucidated for various real-world systems such as the brain \cite{lord2016insights}, protein interaction networks \cite{estrada2018centralities}, ecological communities \cite{grilli2017higher}, and co-authorship networks. 
Combining higher-order interactions with inertia in the Kuramoto model leads to prolonged hysteresis in the synchronization profile \cite{sabhahit2024prolonged}. 
However, this work focuses on large network sizes, where dynamical behaviors align with analytical predictions made in the continuum limit. 

Mean-field studies enable analytical calculations of stable and unstable states, which are pertinent to understanding the origin of synchronization transitions. However, real-world systems are finite in size, causing fluctuations in the order parameter that measures the strength of synchronization. The work of
Daido systematically investigates the finite-size effects of the coupled Kuramoto oscillators model without inertia and having pairwise interactions \cite{HDaido_1987, DAIDO1990, DAIDO199624}.
Suman and Jalan have recently investigated finite-size effects for coupled Kuramoto oscillators with higher-order interactions \cite{suman2024finite}. These investigations for pairwise (1-simplex) and higher-order (2-simplex) interactions were limited for coupled Kuramoto oscillators without inertia.

In this study, we examine the effects of finite size on the dynamics of coupled oscillators through higher-order interactions. We focus on the Kuramoto model with inertia and 2-simplex interactions to present numerical results across various system sizes and inertia values. Although starting with random phases, 2-simplex interactions do not promote synchronization transition in the thermodynamic limit, here, we demonstrate that due to the finite size of the system, a first-order transition to a synchronized state occurs at a finite coupling strength. To elucidate the origin of these finite-size-induced transitions, we provide mean-field analytical calculations for the order parameter, detailing all stable and unstable solutions. This combined numerical and mathematical analysis demonstrates that when fluctuations in the order parameter resulting from the finite size of the networks exceed the unstable state, the system transitions from one stable (incoherent) state to another stable (synchronized) state. Moreover, our simulations suggest that while the finite size aids the system's transition to a synchronized state, an increase in inertia helps restore the results observed in infinite size by shifting the transition point further.

In this work, we will first explain the model used in this study, followed by analytical demonstration of the dynamics of the order parameter with and without inertia. We will then discuss the finite size effects and the role of inertia. Finally, we will compare the numerical results for different system sizes and inertia values with the analytical results.

\paragraph*{\bf{Model:}}
 The coupled Kuramoto model with inertia and  triadic coupling term is given as;
 \begin{equation}
m \ddot {\theta_i} + \dot{\theta_i} =\omega_i+ \frac{K_2}{N^2} \sum\limits_{j,l=1}^{N}\sin(2\theta_j-\theta_l-\theta_i),
\label{model_1d_2}
\end{equation}
where $\theta_i$ and $\omega_i$ indicate the phase angle and intrinsic frequency of $i^{th}$ Kuramoto oscillator, respectively. $K_2$ is overall  coupling strength of triadic interactions and $m$ is the inertia term. $N$ is the total number of coupled oscillators, which we refer to as the size of the system. 
To quantify the extent of synchronization, Kuramoto proposed a complex-valued quantity that is equal to the centroid of the instantaneous phases of all the oscillators in the complex plane as
\begin{equation}
     z_p = r_p e^{i \psi_p} = \frac{1}{N} \sum_{j=1}^N e^{i p \theta_j},
\label{eq_order_parameter}
\end{equation}
where for $p=1$, one can calculate the classical order parameter, i.e., $r_1$, which measures the strength of global synchronization. $r_1=0$ is corresponding to an incoherent state, while $r_1=1$ indicates a completely synchronized state. Eq.~(\ref{model_1d_2}) can be written in the mean-field form using Eq.~(\ref{eq_order_parameter}), explaining that an individual oscillator interacts with a mean force field created by all oscillators in the system. The mean-field equation can be written as
\begin{equation}
    m \ddot {\theta_i} + \dot \theta_i = \omega_i + K_2 r_2 r_{1} \sin(\psi_2 -\psi_1 -\theta_i).
    \label{eq_MF}
\end{equation}
where $r_2$ is calculated using Eq.~(\ref{eq_order_parameter}) for $p=2$. 
\begin{figure}[t]
\begingroup
\begin{tabular}{c c}
\includegraphics[width=0.25\textwidth]{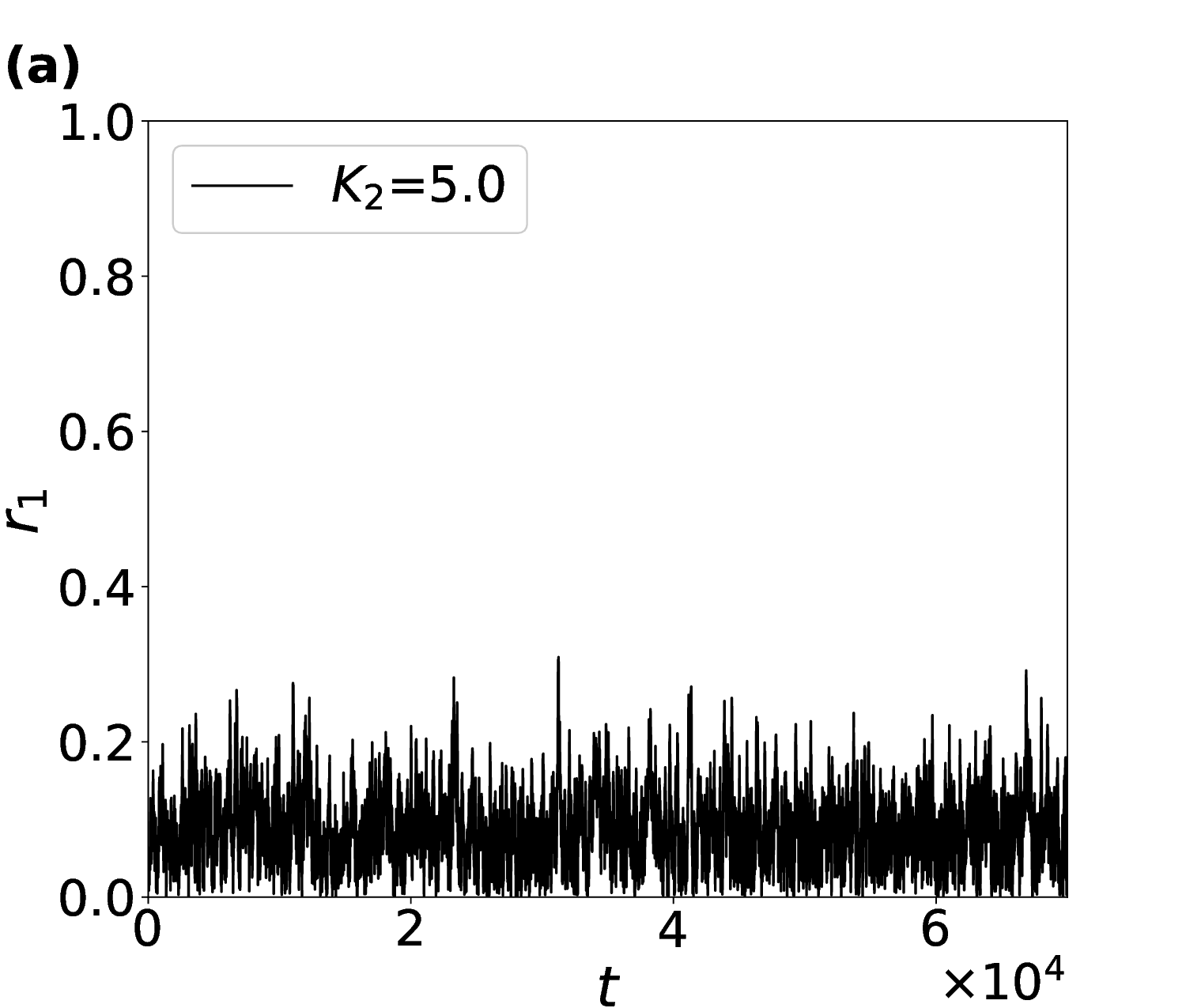}
\includegraphics[width = 0.25\textwidth]{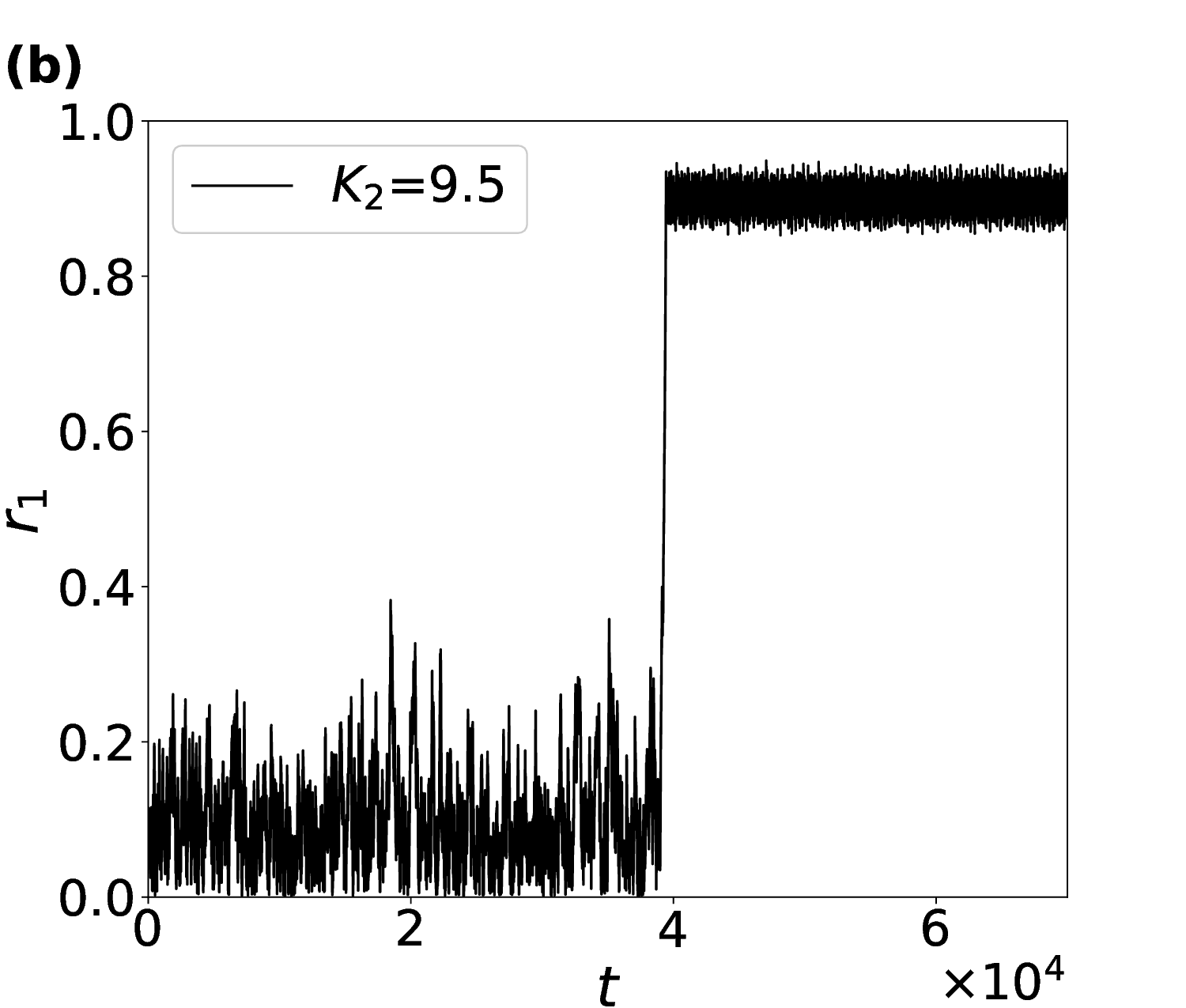}
\\
\includegraphics[width=0.25\textwidth]{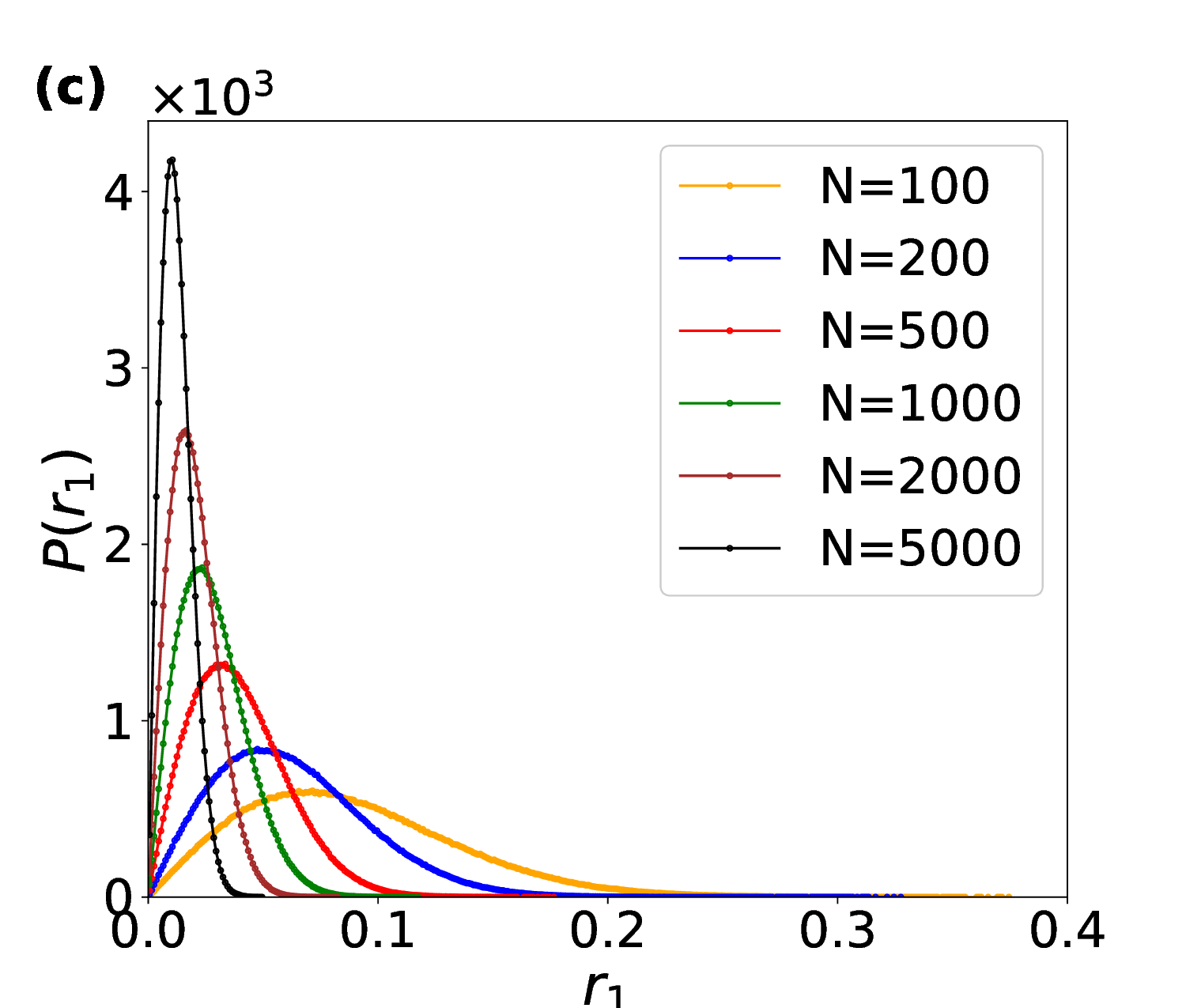}
\includegraphics[width=0.25\textwidth]{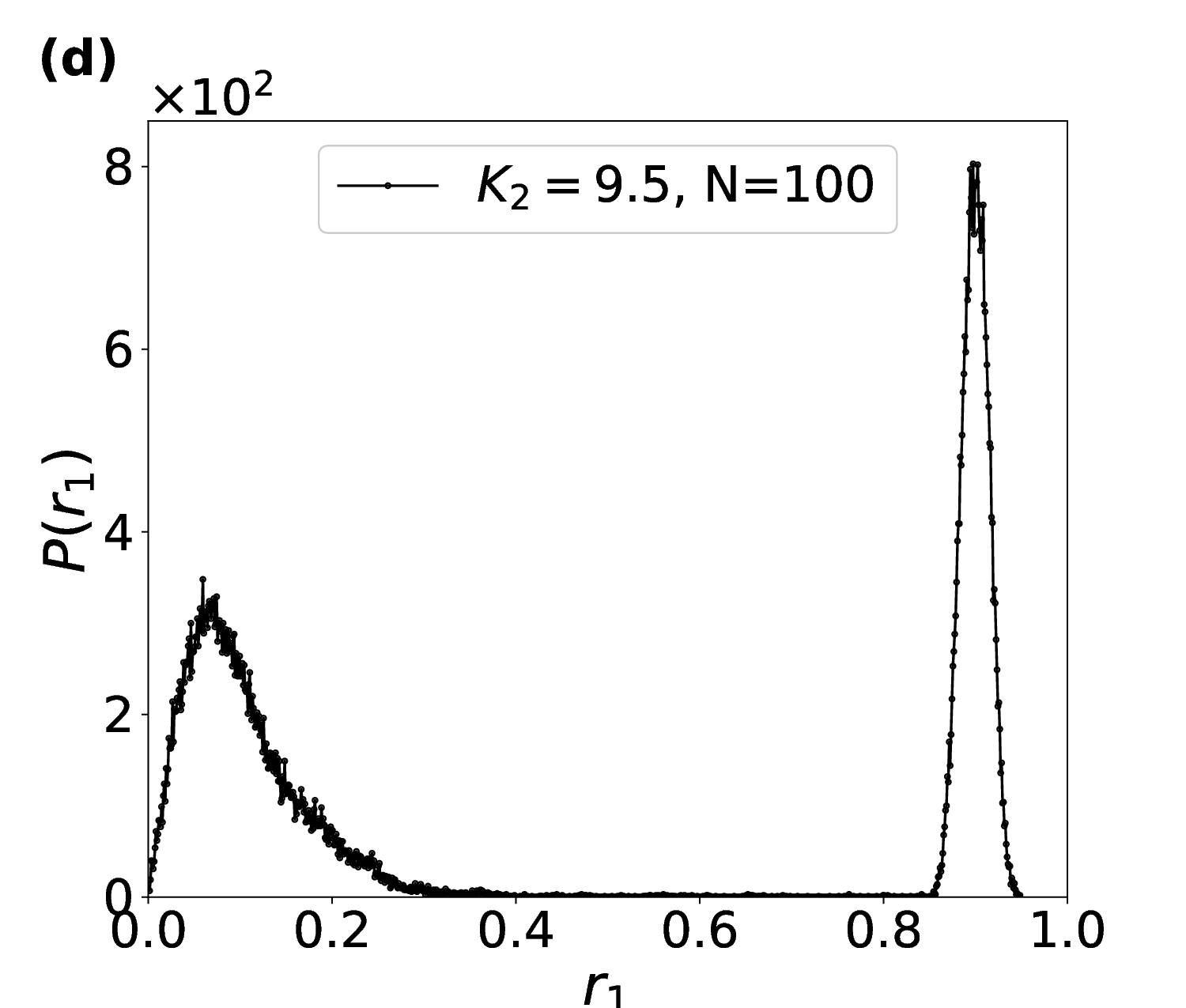}
\end{tabular}
\endgroup
\caption{(a) and (b) plot the time series of $r_1$ for $N=100$. (a) $K_2=5$ corresponding to incoherent state, and (b) $K_2 = 9.5$ where synchronization transition occurs in time. (c) Normalized probability density of $r_1$ for $N = [100, 200, 500, 1000, 2000, 5000]$ for $K_2 = 5.0$ and for $50$ realizations. (d) Normalised probability distribution for $N=100$ at $K_2 = 9.5$. All simulations are for $m=0$.} 
\label{Fig_time_r1_m0}
\end{figure}

\paragraph*{\bf{Analytical Derivation:}}
In steady state, the oscillators synchronize and move with a common mean frequency. 
We consider a Lorentzian frequency distribution for the intrinsic frequency of oscillators given as $g(\omega) = \frac{\Delta}{\pi [(\omega -\omega_0)^2 +\Delta^2]}$, with $\omega_0 = 0$ and the standard deviation is $\Delta=1$.

\begin{figure}[t]
\includegraphics[width=0.43\textwidth]{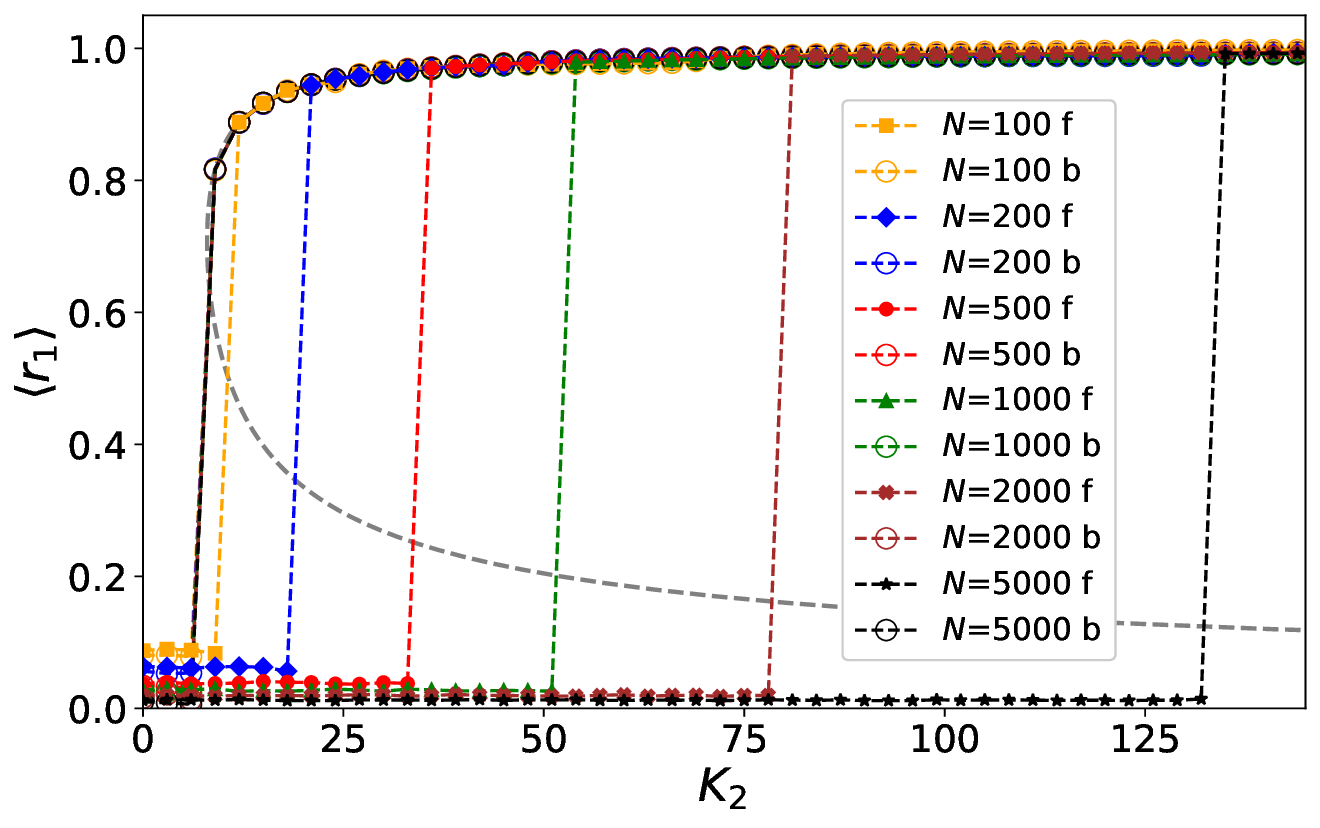}

\caption{Variation of the order
parameter $\langle r_1\rangle_{t\rightarrow\infty}$ as a function of coupling strength $K_2$ for $m=0$. The time average of the order parameter $r_1$ has been plotted for $N=[100, 200, 500, 1000, 2000, 5000]$. The postscript $f$ and $b$ in the legend specify forward and backward direction, respectively. The black {dashed} line represents the analytical prediction of $r_1$  by solving Eq.~\ref{final_analytical_m0} using bisection method, where step size for $r_1$ is $0.0001$ and step size for $K_1$ is $0.001$} 
\label{Fig_r1_k2_m0}
\end{figure}

 For the case of $m \neq 0$, we apply the transformation $\theta_i \rightarrow \theta_i + \Omega t$, with $\Omega$ representing the uniform angular frequency of oscillators that are phase locked. In the rotating frame, $\psi_1$ and $\psi_2$ can be set to zero. Thus, writing Eq.~(\ref{eq_MF}) such as:
\begin{equation}\label{simplify_m_eq}
    m \ddot {\theta_i} = - \dot \theta_i + \omega_i - q \sin(\theta_i),
\end{equation}

\begin{figure}[t]
\begingroup
\begin{tabular}{c c}
\includegraphics[width=0.24\textwidth]{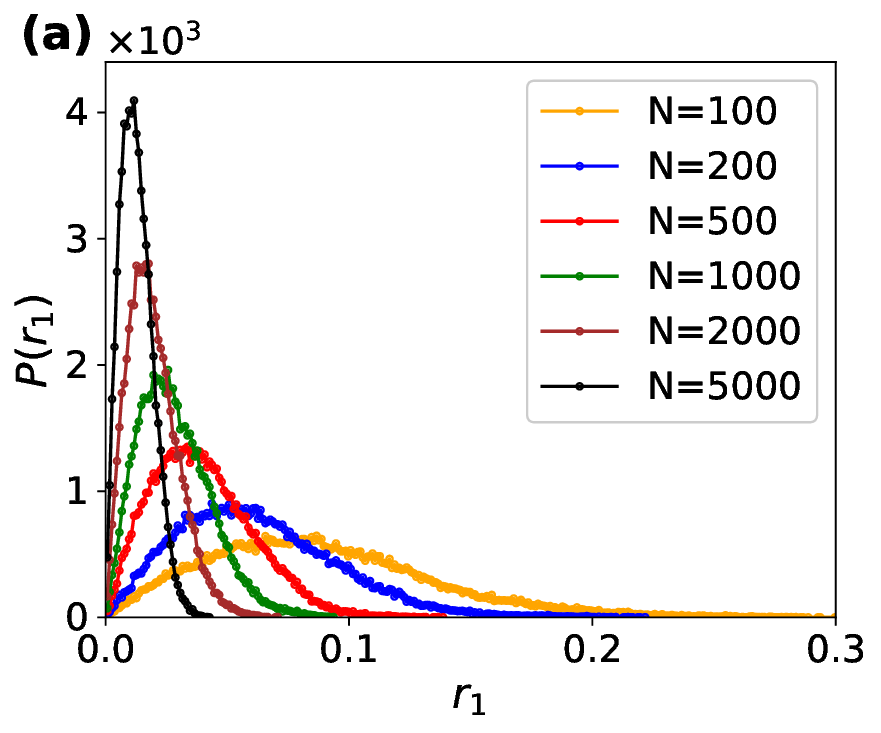}
\includegraphics[width = 0.24\textwidth]{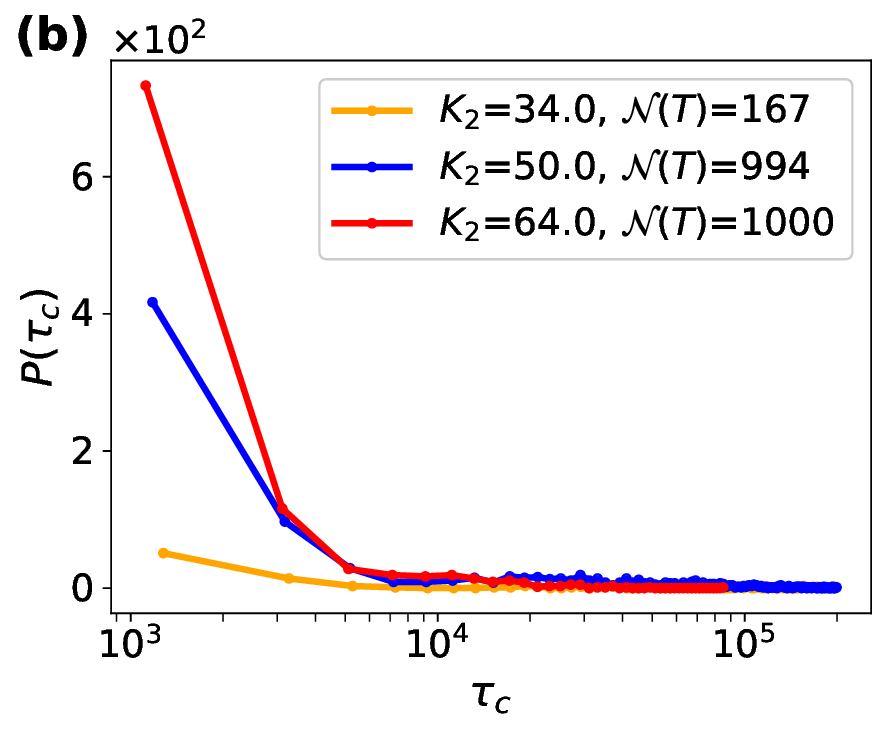}
\end{tabular}
\endgroup
\caption{(a) Normalized probability density of $r_1$ for different values of $N$ for $K_2 = 5.0$ and $m=1$. (b) Distribution of time needed to reach the synchronized state ($\tau_c$) for different values of $K_2$. $\mathcal{N}(T)$ shows the number of times a transition happens for 1000 realizations. All simulations are for $m=1$ and $T=200000$.} 
\label{Fig_time_r1_m1}
\end{figure}

\begin{figure}[t!]
\includegraphics[width=0.43\textwidth]{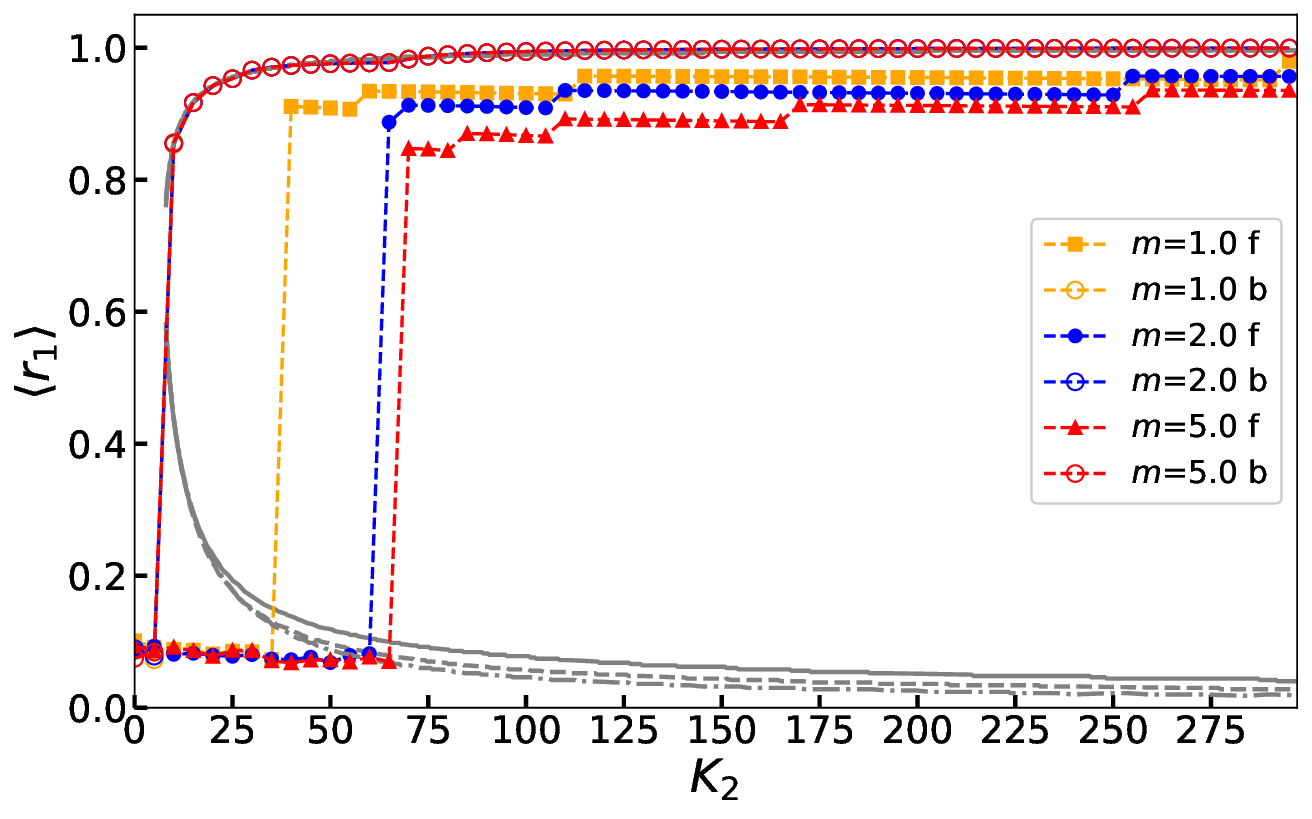}
\caption{The variation of mean order
parameter $\langle r_1\rangle_{t\rightarrow\infty}$ with $K_2$ for $N=100$ and $m=1,2,5$. The postscript $b$ and $f$ in the legend specify forward and backward, respectively. The black lines represent analytical results using Eqs.~(\ref{r_locked}) and (\ref{r_drift}). As shown in the figure, the increase in $m$ notably shifts the transition point, $K_{2c}$, for networks of the same size.}\label{Fig_r1_k2_m_N100}
\end{figure}
\begin{figure*}[t]
\begingroup
\begin{tabular}{c c}
\includegraphics[width=0.31\textwidth]{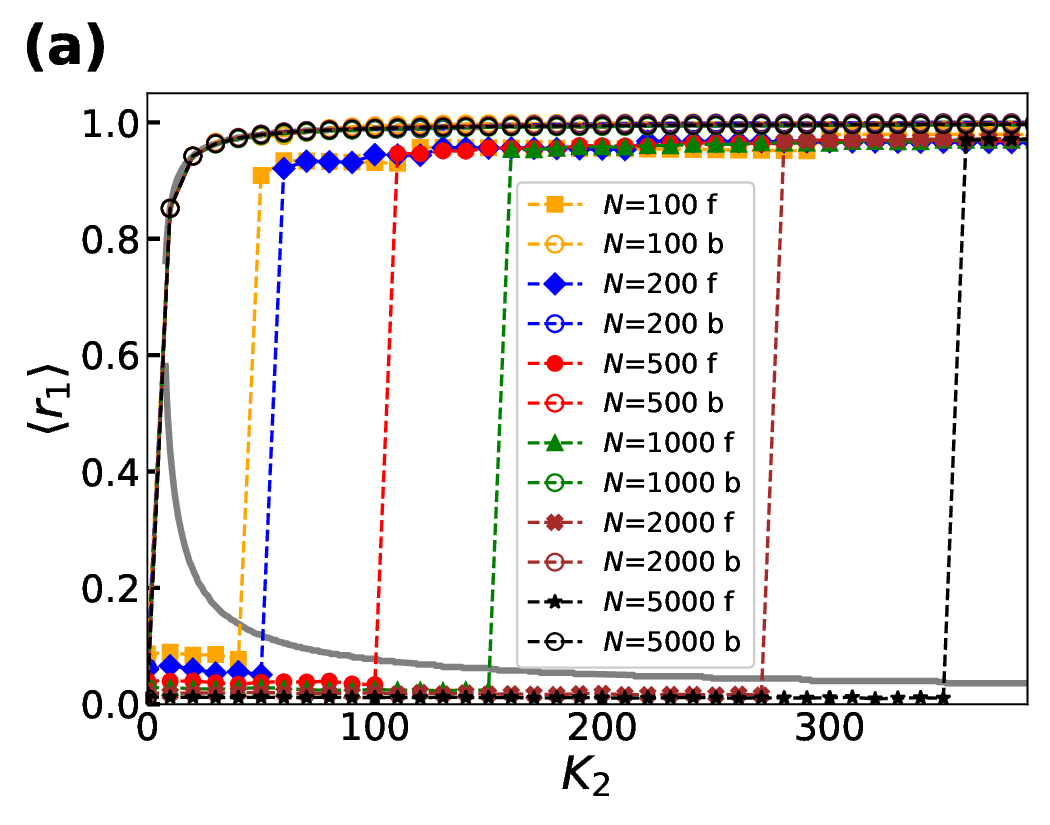}
\includegraphics[width=0.31\textwidth]{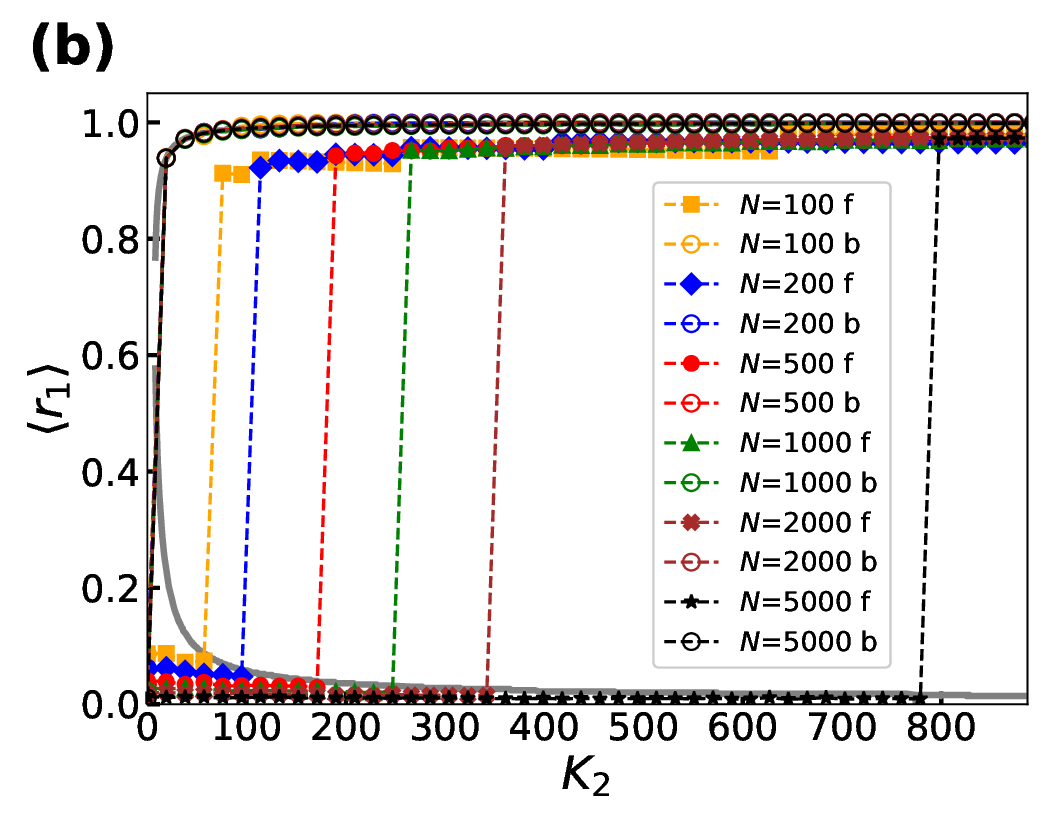}
\includegraphics[width=0.31\textwidth]{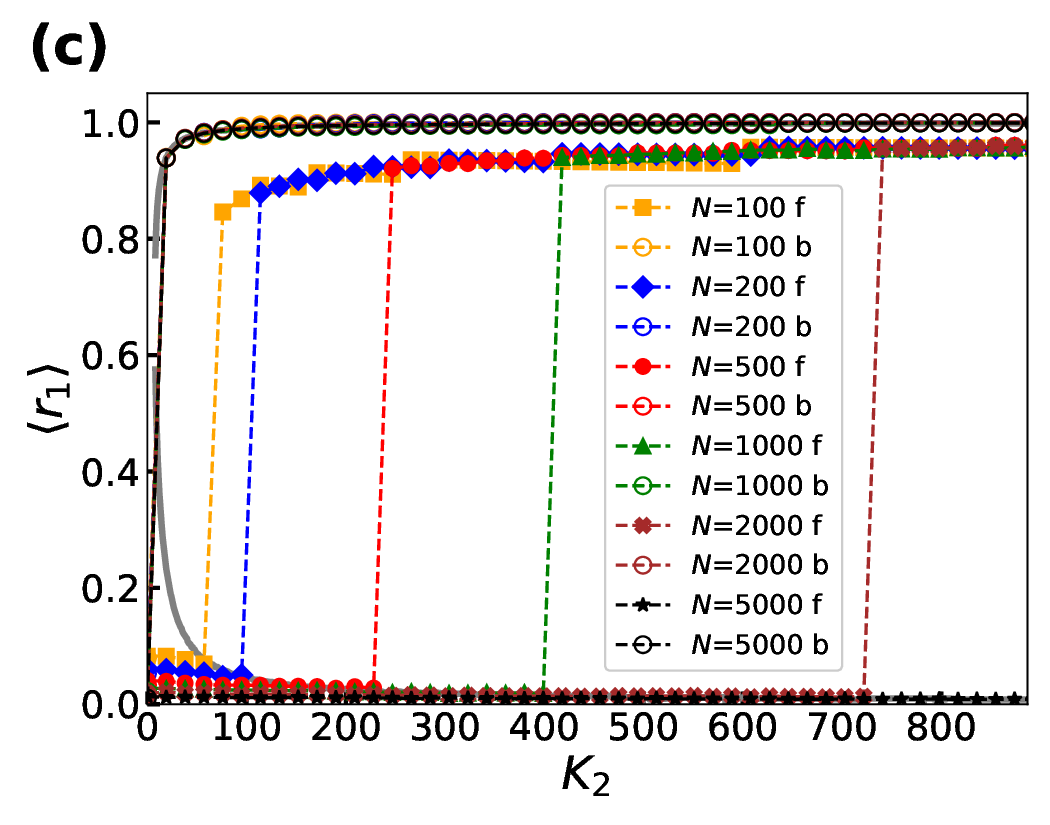}
\end{tabular}
\endgroup
\caption{Variation of the mean order
parameter $\langle r_1\rangle$ with $K_2$ for different values of $N$ and $m$. (a) $m=1$, (b) $m=2$, (c) $m=5$. The postscript $b$ and $f$ in the legend specify forward and backward direction, respectively. The black lines in (a)-(c) represent analytical results (see Eqs.~(\ref{r_locked}) and (\ref{r_drift})). Comparing the results from (a) to (c) indicates that increasing $m$ for networks of the same size significantly raises the transition point. For example, for $N=5000$ and $m=5$ in (c), there is still no transition at $K_2=900$.}
\label{Fig_m_N}
\end{figure*}
\begin{figure}[h]
\includegraphics[width=0.475\textwidth]{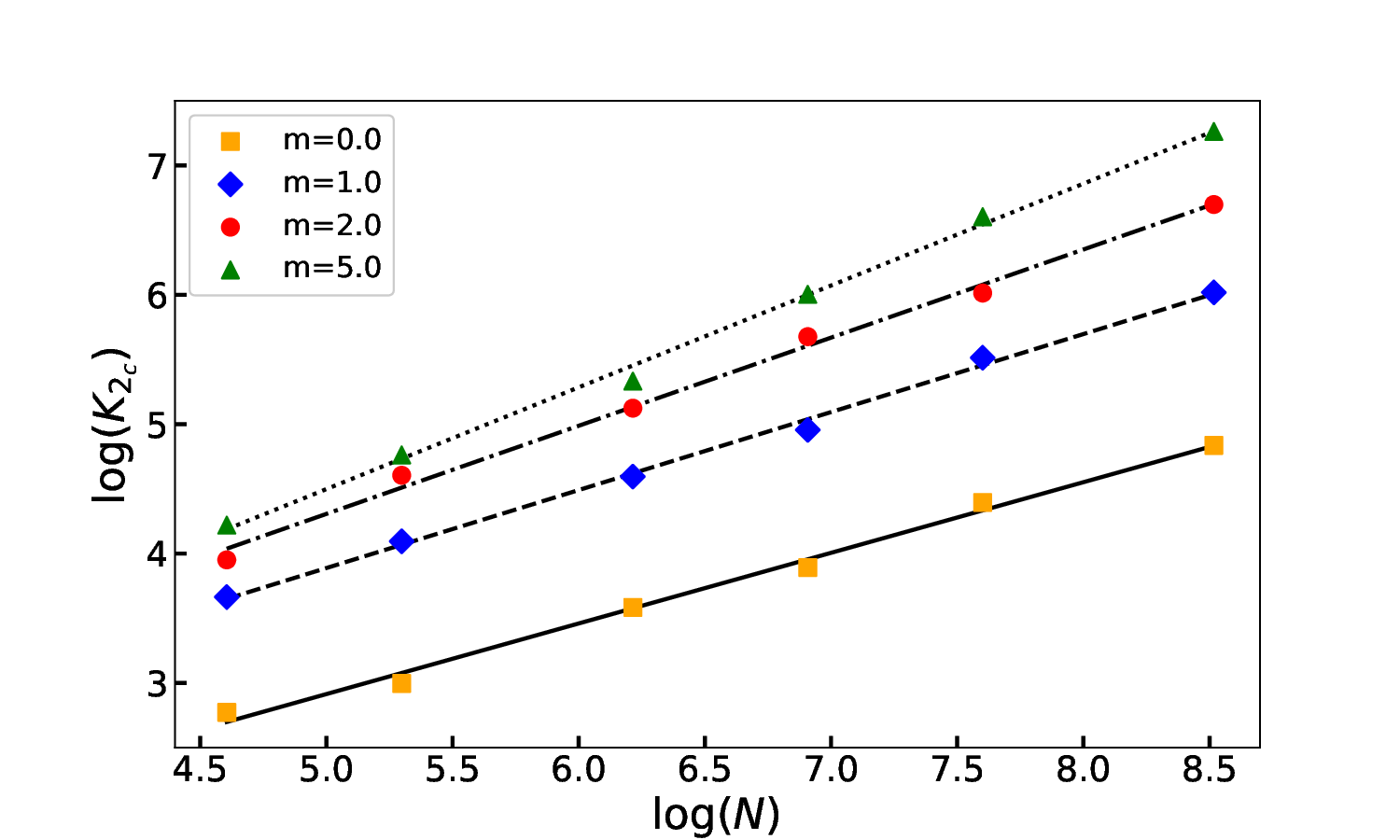}
\caption{Critical coupling value $K_{2_c}$ as a function of network size $N$ for $m=0,1,2,5$. Both axes are presented in logarithmic scale. The relation between $N$ and $K_{2c}$ is $K_{2c} \propto N^\gamma$, where $\gamma$ is $0.55$, $0.58$, $0.67$ and $0.73$ for $m$ equals $0$, $1$, $2$ and $5$, respectively. Note that the results have been normalized with respect to the network size.}
\label{Fig_logplot}
\end{figure}
where $q=K_2 r_2 r_{1}$.  Eq.~(\ref{simplify_m_eq}) is similar to that given in \cite{sabhahit2024prolonged} considering ($K_1=0$). Therefore, following a similar analysis and changing the time scale of the system to $\tau = \sqrt{\frac{q}{m}} t$, we can write Eq.~(\ref{simplify_m_eq}) as 
{{\begin{equation}\label{intertia_simple}\nonumber
    \ddot {\theta} = - \alpha \dot \theta + \beta - \sin(\theta),
    \end{equation}}}
    where $ \alpha = \sqrt{\frac{1}{qm}}$ and $\beta = \frac{\omega}{q}$. The parameter space between $\beta$ and $\alpha$ helps us to find the range of oscillator frequencies that attain the stable fixed point state \cite{gao2018self}. Furthermore, in the steady state, the oscillators population is divided into two groups based on their intrinsic frequencies. One group of oscillators whose frequency is close to the mean frequency, defined as $|\omega|\le q$, locks to the mean phase, while the other group has frequencies that exceed the threshold $|\omega| > q$, drifts around the locked oscillators. The phase coherence of the system can be written as the sum of the contribution of the locked and drifting oscillators.
In continuum limit $N\to\infty$, the general order parameter is defined as $r_p e^{\iota\psi_p}=\int_{-\infty}^{\infty}\int_{-\pi}^{\pi} e^{\iota p\theta}\rho(\theta,\omega)g(\omega)d\theta d\omega$. Following the further analytical derivation in ~\cite{sabhahit2024prolonged}, we write the contribution of locked oscillators as 
{{\begin{equation}\label{r_locked}
    r_p^{l}= q\int_{-\arcsin({\omega/q})}^{\arcsin({\omega/q})} \cos{\theta}\cos(p\theta)g(q\sin\theta)d\theta.
\end{equation}}}
\begin{figure*}[t]
\begingroup
\begin{tabular}{c c}
\includegraphics[width=0.33\textwidth]{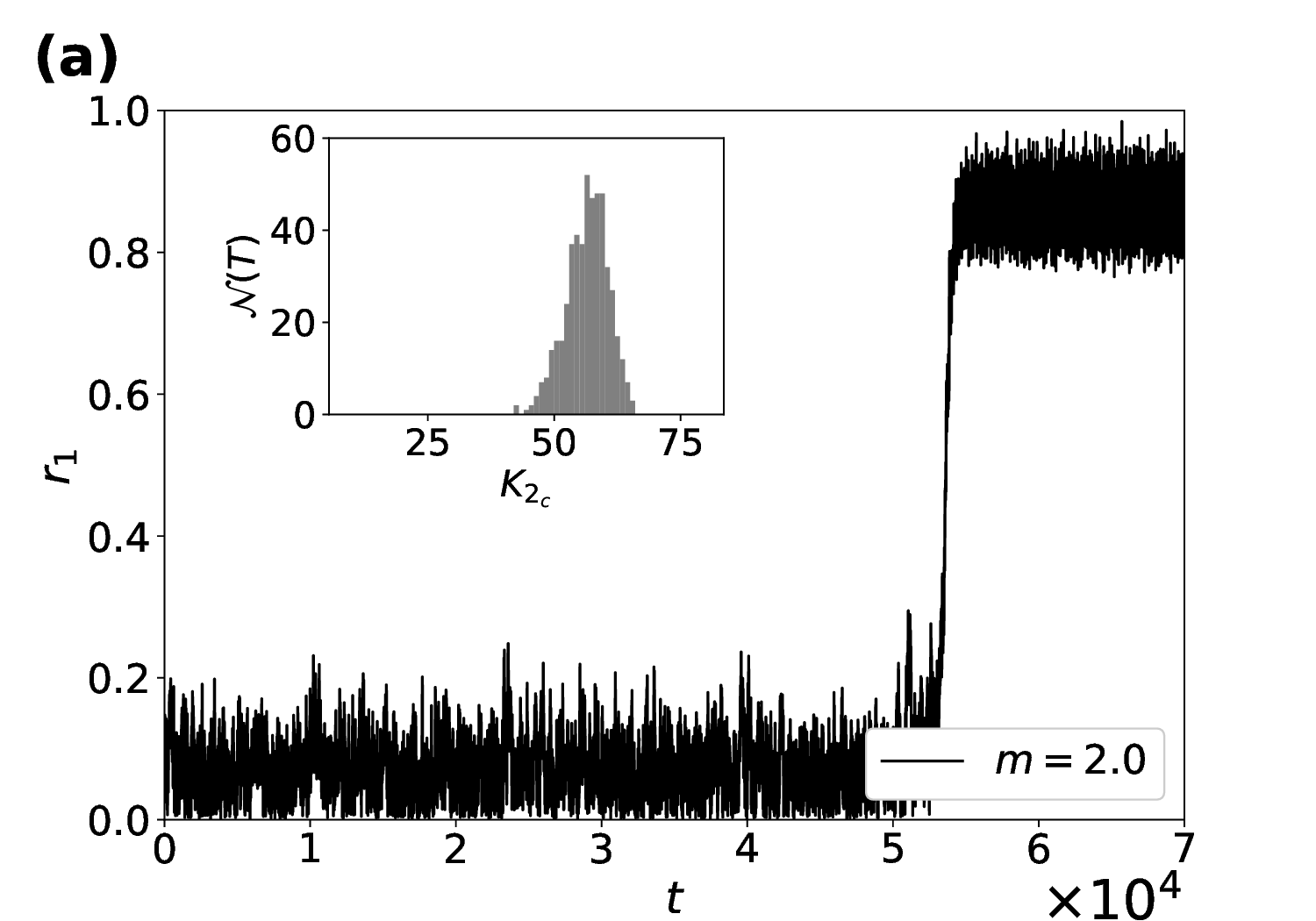}
\includegraphics[width=0.33\textwidth]{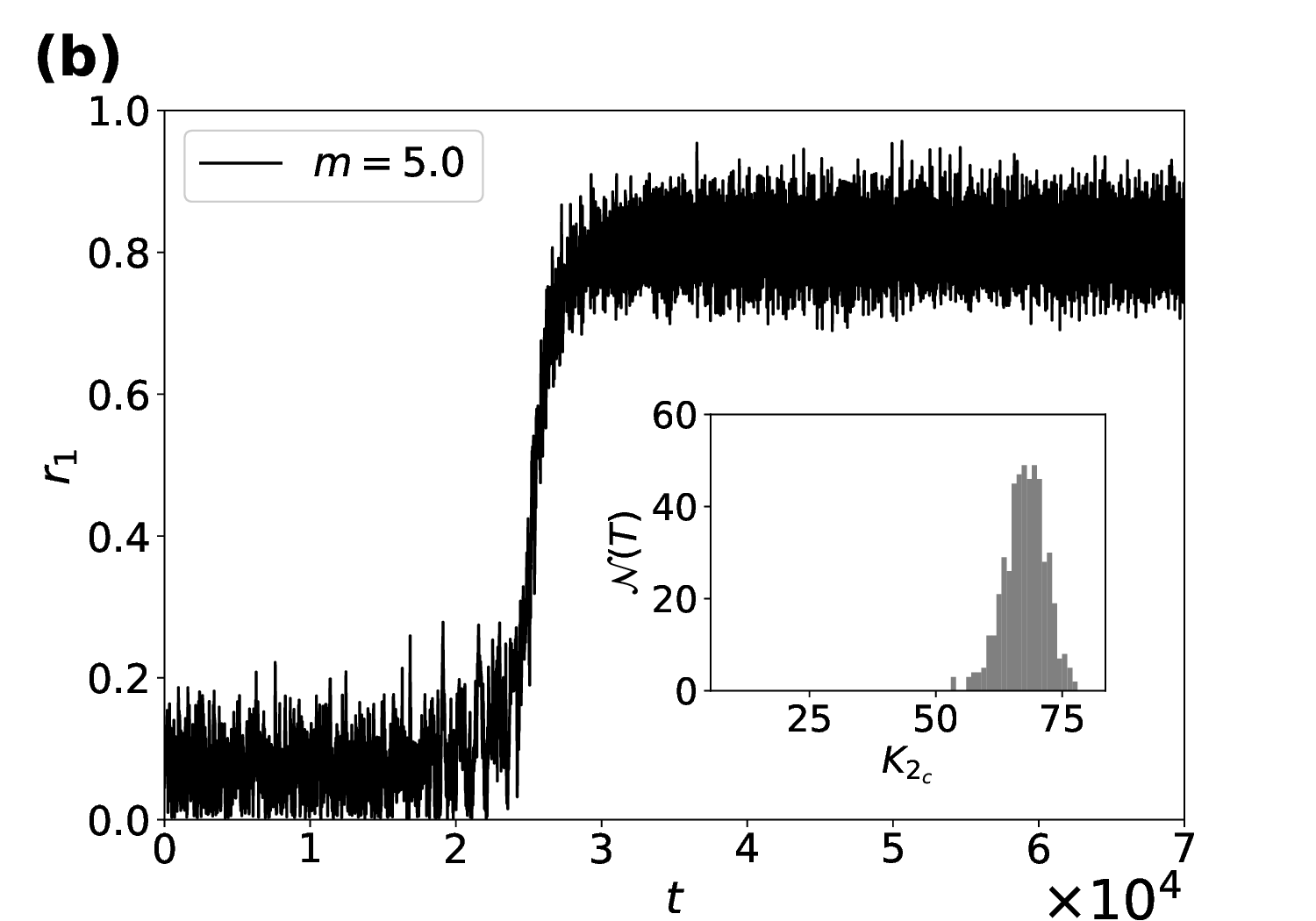}
\includegraphics[width=0.33\textwidth]{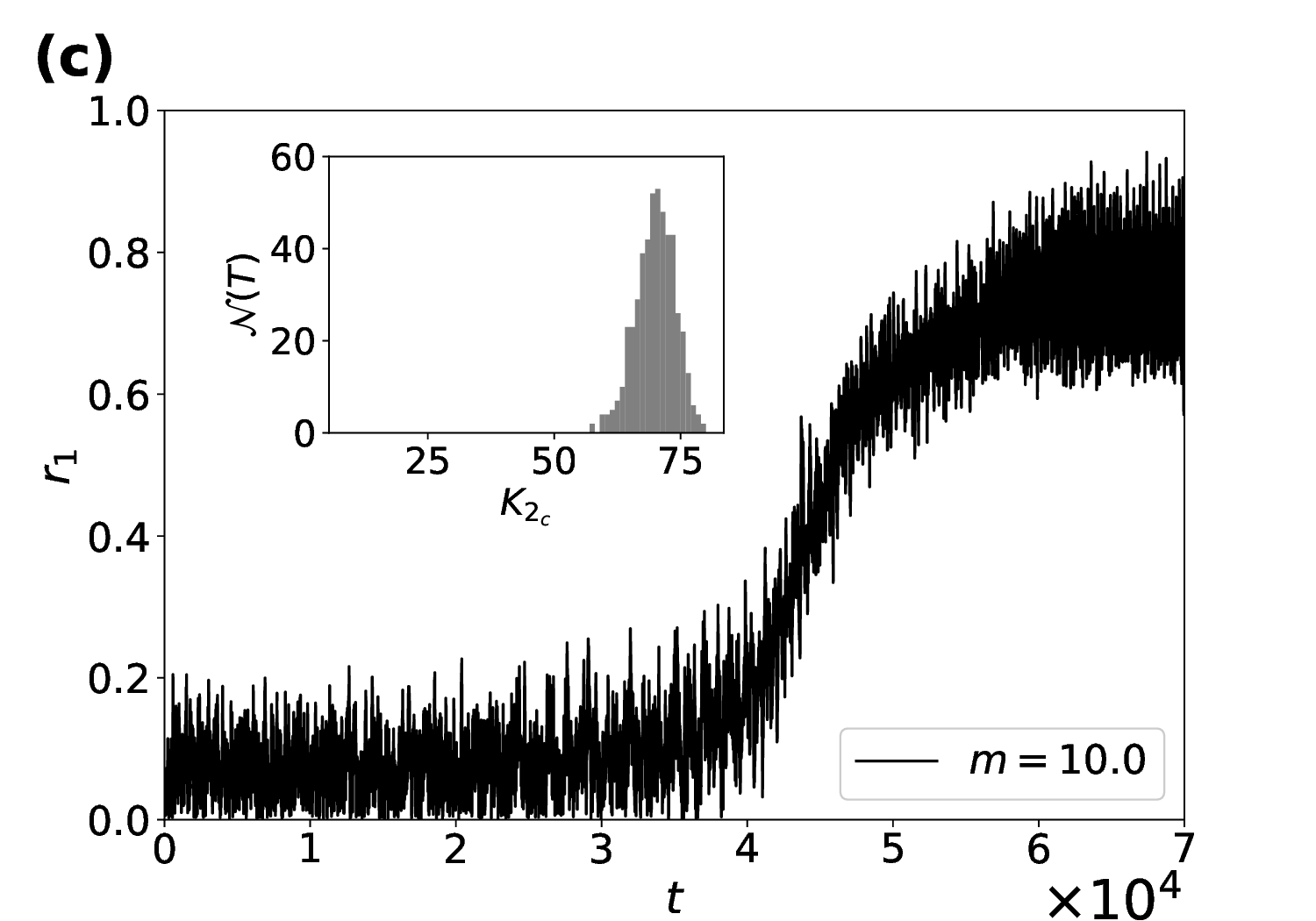}
\end{tabular}
\endgroup
\caption{Time series of $r_1$ for $N=100$ and the first $K_2$ value for which transition from the incoherent to the coherent state occurs. (a) $m=2$ and $K_2=42$, (b) $m=5$ and $K_2=53$, and (c) $m=10$ and $K_2=57$. The inset plots in (a) to (c) show the number of transitions from the incoherent state to the coherent state, $\mathcal{N}(T)$, for different values of $K_2$ for 500 realizations. 
  } 
\label{Fig_m_Fluctuations}
\end{figure*} 
In steady state, where $\dot\theta=\ddot\theta=0$ in Eq.~(\ref{simplify_m_eq}), we obtain $\theta^*=\arcsin({\omega/q})$. Further, the contribution of drifting oscillators is \cite{sabhahit2024prolonged}
\begin{equation}\label{r_drift}
 r_p^{d}=\int_{|\omega|>q}\langle \cos(p\theta) \rangle g(\omega) d\omega,  
\end{equation}
here, for $(p\in 1,2)$, 
\begin{equation}\nonumber
 \begin{split}   
\langle \cos(\theta) \rangle &= \frac{\beta}{\alpha} \left[ \sqrt{\frac{\beta^2}{\alpha^2} - \frac{\alpha^2}{\beta^2 + \alpha^4}} - \frac{\beta}{\alpha} \right]\\
\langle\cos(2\theta)\rangle &= \left[\frac{\beta^2 - \alpha^4}{\beta^2 + \alpha^4}\right] \times 
        \\ & \left[ \frac{2\beta(\beta^2 + \alpha^4)}{\alpha^3}\left(\frac{\beta}{\alpha} - \sqrt{\frac{\beta^2}{\alpha^2} - \frac{\alpha^2}{\beta^2 + \alpha^4}} \right) - 1\right].
\end{split}        
\end{equation}        
        Hence, $\langle r_p\rangle=r_p^{l}+r_p^{d}$ calculated using Eqs.~(\ref{r_locked}) and ~(\ref{r_drift}).
  For the case of $m=0$, we present the analytical calculation in Appendix Eq.~\ref{final_analytical_m0} to provide background for comparing with the $m\neq0$ case.
 Note that Tanaka et al. \cite{TANAKA1997279} calculated the critical point in forward direction for a uniform frequency distribution. Later, Tanaka et al. \cite{PhysRevLett.78.2104tanaka} used the Poincaré–Lindstedt (perturbation) method to derive a closed-form expression for $r_1$ for pairwise interactions for Lorenzian frequency distribution, and by using Melnikov’s method provided a frequency limit $|\omega_p| < (4 \pi)\sqrt{q/m}$ for weak synchronized states.  Here, it is difficult to solve the integral in Eq.~\ref{r_drift} due to the presence of higher-order interactions, which requires simultaneously determining the integral expression for $r_2$. However, we present analytical calculations for $|\omega|< q$ for strong synchronized state.
\\
\paragraph*{\bf{Numerical Results:}}
In a globally coupled system of Kuramoto oscillators (without inertia) on a 2-simplex, in the thermodynamic limits there exists no synchronization transition in the forward direction. When starting from an incoherent state, the stable solution remains at $r_1=0$ for all values of $K_2$. Whereas, starting with a synchronized solution, that is, $r_1=1$, upon decreasing $K_2$ adiabatically, the system manifests a first-order jump to the incoherent state at a backward critical coupling strength \cite{skardal2020higher,Kachhvah_2022}. When inertia is included in the coupled Kuramoto oscillators on a 2-simplex, 
phenomenon remains unchanged; there exists no synchronization transition in the forward direction, while in the backward direction desynchronization occurs accompanied by a saddle-node bifurcation \cite{PhysRevE.109.024212narayan}. 
In conclusion, in the thermodynamic limit, the Kuramoto model - both with and without inertia on a 2-simplex - exhibits a single stable state, $r_1=0$, in part of the $K_2$ region. Beyond the critical backward transition, two stable states, {\color{red}$r_1=0$ and $r_1=1$,} can be achieved depending on the initial conditions. These two stable states are separated by an intermediate unstable state \cite{skardal2020higher,Kachhvah_2022}. 

In this article, we show that for coupled Kuramoto oscillators on finite-sized networks, the stationary states of $r_1$ differ from those predicted through analytical calculations. The value of $r_1$ fluctuates due to the finite size of the underlying globally coupled networks. These fluctuations strongly depend on the size of the system with $\mathcal{O}(1/\sqrt{N})$.

 We used the Runge-Kutta 4 method to simulate Eq.~(\ref{eq_MF}) for time $t$. Figs.~\ref{Fig_time_r1_m0} (a) and (b) plot time series of $r_1$ under two different conditions; $K_2=5$ (corresponding to incoherent state) and $K_{2}=9.5$ (corresponding to the bi-stable region).The figures clearly illustrate the fluctuations of $r_1$ over time due to the finite size effect. Increasing the value of $K_2$ enables the system to transition from an incoherent state to a coherent one.
 In the finite-time scale, $r_1$ forms a time-dependent probability distribution $P(r_1)$ illustrated by Fig.~\ref{Fig_time_r1_m0} (c) obtained by normalizing the histogram of data points of $r_1(t)$ in time for various values of $N$.
 
As reflected from Fig.~\ref{Fig_time_r1_m0} (c), one gets a smaller FWHM (full-width at half-maximum) for a larger number of oscillators. 
 FWHM is inversely proportional to the function of the number of oscillators present in the system (Fig.~\ref{Fig_time_r1_m0} (c)) \cite{suman2024finite}.
 The incoherent state has a finite probability of crossing the unstable state due to fluctuations in $r_1$. Fig.~\ref{Fig_time_r1_m0} (d) illustrates a bimodal distribution, with one peak corresponding to the incoherent state and the other to the synchronized state  for $N=100$ with particular parameter values $K_2=9.5$ and $m=0$. Once the system crosses the unstable state, it settles into another stable (synchronized) state. Consequently, the critical transition points for finite-size systems differ from the analytically determined system for $N \rightarrow \infty$ due to fluctuations in order parameter $r_1$. 
A system with a finite number of oscillators transitions from an unstable state to a synchronized state, as shown in Figs.~\ref{Fig_time_r1_m0} (b) and (d) for $N=100$. This transition occurs due to fluctuations that are inversely proportional to the square root of the system size $N$.

Fig.~\ref{Fig_r1_k2_m0} shows the forward and backward transition of $\langle r_1\rangle_{t\rightarrow\infty}$ as a function of $K_2$ for different network sizes $N = [100, 200, 500, 1000, 2000, 5000]$. Although analytical predictions in the thermodynamic limit do not allow a forward-direction synchronization as $r_1=0$ remains a stable solution for all the values of $K_2$ (Eq.~\ref{final_analytical_m0}), such a transition exists for finite-sized networks arising due to fluctuations in $r_1$ which can be large enough to surpass the unstable branch (dashed black line in Fig. \ref{Fig_r1_k2_m0}) and settling to another stable branch denoted by $r_1=1$ solution (solid black line in Fig. \ref{Fig_r1_k2_m0}). As illustrated in Fig.~\ref{Fig_time_r1_m0} (c), as the network size increases, the FWHM and fluctuations in $r_1$ decrease, and therefore it takes a larger coupling value to cross the unstable state, hence shifting the forward critical coupling $K_{2c}$ towards higher coupling values. The $K_{2c}$ extends to infinity in the thermodynamic limit. However, the backward transition point remains the same regardless of  $N$(Fig.~\ref{Fig_r1_k2_m0}).

Fig.~\ref{Fig_time_r1_m1} (a), similar to Fig.~\ref{Fig_time_r1_m0} (c), presents the time-dependent probability distribution for different system sizes with $m=1$. As observed, the FWHM continues to decrease with an increasing number of oscillators, consistent with the $m=0$ case. Fig.~\ref{Fig_time_r1_m1} (b) illustrates the probability density function (PDF) of the transition times for 1000 realizations across different $K_2$ values. It is evident that for larger $K_2$, the transition from the incoherent state to the coherent state occurs in a shorter time for more realizations.

{Figs.~\ref{Fig_r1_k2_m_N100} and \ref{Fig_m_N} (a)-(c) indicate that the critical coupling strength for 
the backward transition point does not experience any visible change with a change in system size $N$ or inertia $m$. The reason for this is that once the system enters the synchronized state, substantial fluctuations are needed to traverse the unstable state, as it is far from the stable one, i.e., $r_1=1$ (see the black lines in Figs.~\ref{Fig_r1_k2_m0} and \ref{Fig_r1_k2_m_N100}). Therefore, starting with a synchronized state in the backward direction, the system does not depict a transition to an incoherent state due to the finite size effect.} However, the forward critical coupling value varies according to the size of the system and inertia. 
The higher the value of $N$, the smaller the fluctuations in $r_1$, which results in a higher value of $K_{2c}$ for the occurrence of transition to synchronization. Fig.~\ref{Fig_m_N} shows that an increase in $m$ also shifts $K_{2c}$ to the higher values.  Or we can say that an increase in $m$ makes it more difficult to achieve synchronization. 

Fig.\ref{Fig_logplot} summarizes the discussions on the dependence of $K_{2c}$ on $N$ for different values of $m$ to improve the comparison. 
This logarithmic logarithmic graph illustrates that the critical coupling values vary with $N$ according to a power law, i.e. $K_{2c}\simeq N^\gamma$. Moreover, increasing $m$ shifts the value of $K_{2c}$ to the higher value 
indicating that it is more difficult in terms of the values of $K_2$ to achieve the transition as $m$ increases for the same value of $N$. This figure also indicates that inertia resists the effect
caused by the finite size, as with an increase in $m$ not only the overall $K_{2c}$ shift towards higher values, there exists a slight shift in the slope of the straight line characterizing $log(K_{2c}) \simeq log{N}$.


Finally, we analyze the impact of $m$ on fluctuations in $r_1$ and associated $K_{2c}$. 
From Fig.~\ref{Fig_m_Fluctuations}, it is evident that the higher the values of $m$, the larger the value of $K_2$ to witness the transition to the synchronized state. 
As shown in the insets, the critical coupling strength at which the transition from the incoherent state occurs undergoes a general shift to larger values with an increase in $m$.
One can argue that inertia resists the jump induced by the finite size of the system, as seen also in Fig.~\ref{Fig_m_N}. In Fig.~\ref{Fig_m_Fluctuations}, the $K_2$ values used in the simulations to illustrate the time series of $r_1$ correspond to the smallest value at which the transition to synchronization occurs, based on the $500$ realizations for each $m$.  
\\

\paragraph*{\bf{Conclusion:}}

This article investigates the impact of finite size on the phase synchronization of Kuramoto oscillators with inertia and 2-simplex coupling. Although analytical calculations do not suggest a synchronization transition in the thermodynamic limit, we observe that finite-sized networks can experience an abrupt synchronization jump.

This synchronization transition is driven by fluctuations in the order parameter. For smaller networks, these fluctuations can be significant enough to cross the unstable orbit in the bistable region. Starting from a random initial phase condition, the system transitions from an incoherent stable state to a synchronized stable state as soon as fluctuations in $r_1$ become large enough to cross the unstable orbit.

As the size of the system increases, the fluctuations in $r_1$ decrease. Consequently, stronger coupling strengths are generally required to achieve this synchronization jump, since increasing the coupling strength brings the unstable orbit closer to the $r_1 = 0$ state. Additionally, for networks of the same size, increasing $m$ also necessitates a stronger coupling strength to effect this transition from an incoherent state to a synchronized state. This suggests that inertia acts to resist the effects associated with finite network size.
Future research could explore the incorporation of network architectures that better reflect real-world systems.



\section{Acknowledgments}
SJ and MA acknowledge SERB POWER Grant No.
SPF/2021/000136 and IGSTC WISER award with Grant No. 11-20286-3180-00001, respectively. AS and PR thank UGC for a junior research fellowship, and Government of India for the PMRF with Grant No. PMRF/2023/2103358, respectively. 

\appendix
\section{Analytical solution for $m=0$}
{
For $m=0$, the simplified equation can be analytically found using the Ott-Antonsen approach \cite{10.1063/1.2930766}. In the continuum limit ($N \rightarrow \infty$) the state of the system can be described by a density function $\rho(\theta, \omega, t)$ at time t.
\begin{equation}\label{continuity eq}
   \frac{\partial \rho}{\partial t} + \frac{\partial (\rho \dot{\theta})}{\partial \theta}=0,
\end{equation}
Due to the conservation of the number of oscillators, the density function satisfies the continuity equation Eq.~(\ref{continuity eq}). Further, the periodic nature of the phases ($\theta$) allows us to expand the density function into the Fourier series as follows.
\begin{equation}\label{density func}
    \rho(\theta, \omega, t) = \frac{g(\omega)}{2 \pi} [1+ \sum_{n=1}^{\infty} f_n (\omega,t) e^{i n\theta} + {\sum_{n=1}^{\infty} f_n^* (\omega,t) e^{-i n\theta}}],
\end{equation}
where $f_n (\omega,t)$ being $n^{th}$ Fourier coefficient. Using the Ott-Antonsen ansatz \cite{10.1063/1.2930766}, we define $f_n(\omega,t) = \alpha^n(\omega,t)$, where $|\alpha(\omega,t)| \leq 1$. 
For an over-damped system \textbf{($m\approx 0$),} Eq.~(\ref{eq_MF}) will be,
\begin{equation}\label{meanfield}
    \dot \theta_i = \omega_i + K_2 r_2 r_{1} \sin(\psi_2 -\psi_1 -\theta_i).
\end{equation} 
By incorporating Eqs.~ ~(\ref{density func} and \ref{meanfield}) into the continuity equation, i.e., Eq.~(\ref{continuity eq}) and equating the coefficient of $e^{i\theta}$, we derive the dimensional reduction form of Eq.~(\ref{model_1d_2}) without inertia, as follows:
\begin{equation}\label{OAmanifold}
\begin{aligned}
    \dot \alpha &= -i \omega \alpha + \frac{K_2 r_2 r_1}{2} (e^{-i(\psi_2 - \psi_1)} - \alpha ^2 e^{i(\psi_2 - \psi_1)} ). 
\end{aligned}
\end{equation}
To find the relation between $r_1$ and $\alpha(\omega,t)$, in the continuum limit we have,
\begin{equation}\nonumber
    z_p = \int_{- \infty}^{\infty} \int_{0}^{2\pi} \rho(\theta, \omega, t) e^{ip \theta} d\theta d\omega.
\end{equation}  
To solve the above integral, we insert $\rho(\theta, \omega, t)$  from Eq.~(\ref{density func}). Integration over $\theta$ yields $z_1 = \int_{- \infty}^{\infty} g(\omega) \alpha^*(\omega ,t) d\omega$ and $z_2 = \int_{- \infty}^{\infty} g(\omega) \alpha^{*2} (\omega ,t) d\omega$. Furthermore, for Lorentzian intrinsic frequency distribution, we use Cauchy's residue theorem for the lower or upper half of the complex plane, which results in $r_1 e^{-i\psi_1} = \alpha(\omega_0-i\Delta,t)$, and $r_2 e^{-i\psi_2} = (\alpha(\omega_0-i\Delta,t))^2$ 
 \cite{10.1063/5.0224001Abhishek}. Simplifying Eq.~(\ref{OAmanifold}) by putting the value of $\alpha$, we obtain the temporal evolution of order parameter $r_1$ which is led by the following nonlinear differential equation:
\begin{equation}\label{final_analytical_m0}
    \dot r_1 = -r_1 + \frac{K_2}{2} (r_{1}^3-r_{1}^5).
\end{equation}
We can write Eq.~(\ref{final_analytical_m0}) as $\dot r_1 = f(r_1)$. The solutions of $f(r_{1}) = 0$ yields us all the fixed points of $r_1$.
One trivial solution is $r_1=0$; the other two solutions exist in the range $0 < r_1 \le 1$. Among these three solutions, two are stable, and one is unstable. To find the stable and unstable point we use $\frac{\partial f(r_1)}{\partial r_1}$ less than or greater than zero, respectively. An unstable point pushes the trajectory of $r_1$ away, while a stable point attracts the trajectory.
Therefore, a system of finite size will settle to one of the stable fixed points as $t \to \infty$ unless it starts exactly from the unstable point.}
\medskip
\bibliography{main}

\end{document}